\documentclass[12pt]{iopart}

\usepackage{iopams,epsfig}

\newcommand{\be}{\begin{equation}}
\newcommand{\ee}{\end{equation}}
\newcommand{\bea}{\begin{eqnarray}}
\newcommand{\eea}{\end{eqnarray}}

\def\ov{\over}
\def\le{\left}
\def\ri{\right}

\def\lam{{\lambda}}
\def\Lam{{\Lambda}}

\def\vev#1{\langle#1\rangle}
\def\det{{\rm det}}

\def\NN{{\cal N}}
\def\th{{\theta}}

\def \th{{\theta}}

\def \lam {\lambda}

\def\ep{{\epsilon}}

\def\CC{{\cal C}}

\def\GeV{{\rm GeV}}
\def\fm{{\rm fm}}
\def\MeV{{\rm MeV}}

\def\eeq{\end{equation}}

\begin{document}

%\preprint{MIT-CTP-3815\\ hep-ph/0702210}

\title[]{Heavy ion collisions and AdS/CFT}

\author{Hong Liu}

\address{Center for Theoretical Physics, \\
Massachusetts Institute of Technology, \\
Cambridge, MA 02139, USA}
\ead{hong\_liu@mit.edu}
\begin{abstract}

We review some recent applications of the AdS/CFT correspondence
to heavy ion collisions including a  calculation of the jet
quenching parameter in $\NN=4$ super-Yang-Mills theory and
quarkonium suppression from velocity scaling of the screening
length for a heavy quark-antiquark pair. We also briefly discuss
differences and similarities between QCD and $\NN=4$
Super-Yang-Mills theory.

\end{abstract}

%Uncomment for PACS numbers title message
%\pacs{00.00, 20.00, 42.10}
% Keywords required only for MST, PB, PMB, PM, JOA, JOB?
%\vspace{2pc}
%\noindent{\it Keywords}: Article preparation, IOP journals
% Uncomment for Submitted to journal title message
%\submitto{\JPA}
% Comment out if separate title page not required
%\maketitle

\section{Introduction}

Understanding the implications of data from the Relativistic Heavy
Ion Collider (RHIC) poses qualitatively new
challenges~\cite{RHIC}. Given its large and anisotropic collective
flow and its strong interaction with hard probes, the hot matter
produced in RHIC collisions must be described by QCD in a regime
of strong, and hence nonperturbative, interactions. In this
regime, lattice QCD has to date been the prime calculational tool.
However, understanding collective flow, jet quenching and other
hard probes requires real-time dynamics, on which information from
lattice QCD is at present both scarce and indirect. Complementary
methods for real-time strong coupling calculations at finite
temperature are therefore desirable.

For a class of non-abelian thermal gauge field theories, the
AdS/CFT conjecture provides such an alternative~\cite{AdS/CFT}. It
maps nonperturbative problems at strong coupling onto calculable
problems of classical gravity in a five -dimensional anti-de
Sitter (AdS$_5$) black hole spacetime. Although the AdS/CFT
correspondence is not directly applicable to QCD, one expects
results obtained from closely related non-abelian gauge theories
should shed qualitative (or even quantitative) insights into
analogous questions in QCD. A beautiful example is the
universality of shear viscosity in various gauge theories with a
gravity dual~\cite{Policastro:2001yc} and its numerical
coincidence with estimates from comparing RHIC data with
hydrodynamical model analyses~\cite{Teaney:2003kp}.

Here we give a short overview of two recent AdS/CFT calculations
of relevance to heavy ion collisions: (i) the jet quenching
parameter which controls the description of medium-induced energy
loss for relativistic partons in QCD~\cite{Liu:2006ug,Liu:2006he};
(ii) velocity-induced quarkonium
suppression~\cite{Liu:2006nn,Liu:2006he}. %Other related work
%includes~\cite{others}.,Kovtun:2003wp,Buchel:2003tz

\section{Jet quenching and AdS/CFT}

A high energy parton moving in a QCD quark-gluon plasma will lose
energy from interaction with the medium. Medium-induced gluon
radiation has been argued to be the dominant mechanism behind jet
quenching at RHIC (for reviews see~\cite{jetquenchrev}), where the
high energy partons whose energy loss is observed in the data have
transverse momenta of at most about 20 GeV~\cite{RHIC}. In the
Baier-Dokshitzer-Mueller-Peigne-Schiff~\cite{Baier:1996sk}
calculation of the medium-modified splitting processes $q \to q\,
g$, the quark-gluon radiation vertex is treated perturbatively.
 However, rescatterings
of the radiated gluon and the initial quark with the medium are
controlled by $\alpha_s(T)$ and cannot be treated perturbatively
in a strongly interacting quark-gluon plasma. In the high energy
limit, at order $O(1/E)$ with $E$ the energy of the initial quark,
these non-perturbative effects are captured by a single parameter
$\hat q$, which can heuristically be understood as the transverse
momentum square transfer from the medium to the {\it light-like}
initial quark (or radiated gluon) per unit distance. In a heavy
ion collision, $\hat q$ decreases as the hot fluid expands and
cools. The time-averaged $\hat q$ determined in comparison with
RHIC data is found to be around
5-15~GeV$^2/$fm~\cite{Eskola:2004cr,Dainese:2004te}.

A weak-coupling calculation of $\hat q$ in a static medium yields
(up to a logarithm)~\cite{Baier:1996sk,Baier:2002tc,Baier:2006fr}
\begin{equation}
\hat q_{\rm weak-coupling} = \frac{8 \zeta(3)}{\pi} \alpha_s^2 N^2
T^3 \label{2.16}
\end{equation}
if $N$, the number of colors, is large. However, given
$\alpha_s(T)$ at RHIC temperature is not small, a weak coupling
calculation is not under control. Taking $\alpha_s=1/2$ for
temperatures not far above the QCD phase transition and $N=3$, one
finds from (\ref{2.16}) that
 $\hat q_{\rm weak-coupling} \approx 0.94 \; %, 2.24, 4.37\;
 \GeV^2/\fm$
 %\quad {\rm for} \quad   T= 300, 400, 500 \, \MeV\,
 %\ee
for $T=300 \MeV$ (which is roughly the temperature of RHIC at $t=1
\fm$), smaller than the experimental estimate by at least a factor
of 5. There is thus strong motivation to calculate $\hat q$
without assuming weak coupling.

\begin{figure}[t]
%\FIGURE[t]{
%\vskip-0.15in
% \hfill\hskip-0.6in
\includegraphics[scale=0.4,angle=0]{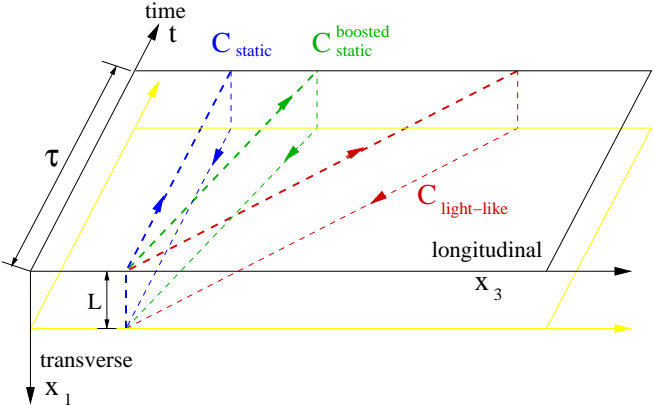}
%\hskip-0.1in
%\hfill
\caption{ Schematic illustration of the shape of Wilson loops
${\cal C}$, corresponding to a $q\bar{q}$ dipole of size $L$,
oriented along the $x_1$-direction, which is  (i) at rest with
respect to the medium (${\cal C}_{\rm static}$), (ii) moving with
some finite velocity $v$ along the longitudinal $x_3$-direction
(${\cal C}_{\rm static}^{\rm boosted}$), or (iii) moving with the
velocity of light along the $x_3$-direction~(${\cal C}_{\rm
light-like}$).} \label{fig0}
%}%
\end{figure}

In~\cite{Liu:2006ug,Wiedemann:2005gm} (see also~\cite{Liu:2006he}
and references therein), a non-perturbative definition of $\hat q$
was given in terms of the short distance limit of the thermal
expectation value of a light-like Wilson loop in the adjoint
representation
  \begin{eqnarray}
\langle{W^A({\cal C}_{\rm light-like})}\rangle
    &\approx& \exp\left[ - \frac{1}{4\sqrt{2}}  \hat{q}\, L^-\, L^2
    \right]\,, \qquad
 L^- \gg {1 \ov T} \gg L
\label{eq5}
 \end{eqnarray}
where the contour ${\cal C}_{\rm light-like}$ is a rectangle with
large extension $L^-$ in the $x^-$-direction and small extension
$L$ in a transverse direction (see fig~\ref{fig0}). At a heuristic
level, the two long sides of the Wilson loop can be understood to
arise from the eikonal phase of the radiated gluon moving in the
medium (which is a lightlike Wilson line along the gluon
trajectory) and its complex conjugate. The transverse separation
$L$ is conjugate to the transverse momentum ${\bf k}_{\perp}$ of
the emitted gluon. In~(\ref{eq5}) we again consider a static
medium and $\hat q$ is constant.

While it is currently not known how to directly
compute~(\ref{eq5}) for QCD in a strong coupling regime, it is of
interest to compute~(\ref{eq5}) in other non-Abelian gauge
theories to extract qualitative information such as how $\hat q$
depends on the number of degrees of freedom, its coupling constant
dependence and so on. At a quantitative level, it would be
interesting to know whether other theories exist which can give
rise to a $\hat q$ as large as the experimental estimate.

For $\NN=4$ Super-Yang-Mills (SYM) theory with gauge group $SU(N)$
in the limit of large $N$ and large 't Hooft coupling $\lam =
g_{SYM}^2 N$, the Wilson loop (\ref{eq5}) can be calculated using
the AdS/CFT correspondence. ${\cal N}=4$ SYM is a supersymmetric
gauge theory with one gauge field $A_\mu$, six massless scalar
fields $X^I, I=1,2,\cdots 6$ and four massless Weyl fermionic
fields $\chi_i$, all transforming in the adjoint representation of
the gauge group. The theory is conformally invariant and is
specified by two parameters: the rank of the gauge group $N$ and
the 't Hooft coupling $\lambda = g_{SYM}^2\, N $. (The coupling
constant does not run.) In the large $N$ and large $\lam$ limit,
the thermal expectation value of a Wilson loop operator $\vev{W
(\CC)}$ in $\NN=4$ SYM theory is given
by~\cite{Rey:1998ik,Rey:1998bq}
  \begin{equation}
\langle{W({\cal C})}\rangle = \exp\left[ i\,  S ({\cal C})
\right] \, ,\label{3.7}
 \end{equation}
where $S(\CC)$ is the classical extremal action of a string in a
five dimensional anti-de Sitter (AdS$_5$) black hole geometry,
with the boundary of the string world sheet ending on the curve
${\cal C}$ lying in the boundary of the black hole spacetime. The
string worldsheet can be considered as the spacetime trajectory of
an open string connecting the quark and antiquark which are
running along the loop $\CC$. The open string ``lives'' in a
$(4+1)$-d AdS$_5$ black hole spacetime with our ($3+1$)-d
Minkowski spacetime as its boundary.

For a light-like Wilson loop~(\ref{eq5}), the extremal string
worldsheet is spacelike and $S(\CC)$ is pure imaginary. Thus the
exponent on the right hand of (\ref{3.7}) is real as in that of
(\ref{eq5}) and one finds that $\hat q$ is given
by~\cite{Liu:2006ug}
 \be
 \hat{q}_{SYM} = \frac{\pi^{3/2} \Gamma\left(\frac{3}{4}\right)}
 { \Gamma\left(\frac{5}{4}\right)}
    \sqrt{\lambda}\, T^3 \approx 26.69 \sqrt{\alpha_{\rm SYM}N} \,T^3\,.
    \label{qhat}
 \ee
Note that $T^3$ behavior in (\ref{qhat}) can be determined by
dimensional analysis since $\NN=4$ SYM is conformal. The
$\sqrt{\lam}$ dependence is non-trivial and is a consequence of
strong coupling.
 Taking $N=3$ and
$\alpha_{\rm SYM}=\frac{1}{2}$, thinking $\alpha_{\rm
QCD}=\frac{1}{2}$ for temperatures not far above the QCD phase
transition, one finds that
  \bea
 \hat q_{\rm SYM} & =   4.5 %, 10.6, 20.7  \;
  \; \GeV^2/\fm \qquad {\rm for} \qquad  & T= 300  %, 400, 500 \,
   \, \MeV\,.
       %  & = 10.6 \;  \GeV^2/\fm \qquad & T= 400 \, \MeV \\
       %& = 20.7 \; \GeV^2/\fm \qquad & T= 500 \, \MeV
 \label{numV}
 \eea
It is both surprising and amusing that the $\NN=4$ SYM answer is
rather close to the experimental estimate mentioned earlier. %In the
%last section we will discuss whether the agreement is meaningful.

One can also use AdS/CFT to evaluate $\hat q$ for other
non-abelian gauge theories with a supergravity
dual~\cite{Buchel:2006bv}. The value of $\hat q$ is not universal
among different theories. Nevertheless, it appears that $\hat q$
can be considered as a measure of the number of degrees of freedom
of a theory at an energy scale $T$~\cite{Liu:2006he}. For example,
for {\it any} conformal field theory which is dual to a type IIB
string theory on AdS$_5 \times M_5$ where $M_5$ is a 5-d Einstein
manifold, one finds that
 \be
 \frac{ \hat q_{CFT}}{\hat q_{\NN=4}}
 = \sqrt{\frac{ s_{CFT}}{ s_{\NN=4 }}} \ ,
 \label{fact}
 \ee
 in the limit of
large $N$ and large 't Hooft coupling $\lam$. Given that QCD at a
temperature of a few $T_c$ appears to be rather close to being
conformal, it is tempting to conjecture that~\cite{Liu:2006he}
 \be \frac{ \hat q_{QCD}}{\hat q_{\NN=4}}
\sim \sqrt{\frac{ s_{QCD}}{ s_{\NN=4 }}} = \sqrt{\frac{47.5}{120}}
\simeq 0.63 \label{conjecture}
 \ee
as an estimate of the effect of the difference between the number
of degrees of freedom in the two theories on $\hat q$.

In a relativistic heavy ion collision, the medium itself develops
strong collective flow, meaning that the hard parton is traversing
a moving medium --- it feels a wind. Thus to compare with the
experimental estimate we should include the effects of the wind on
$\hat q$. The behavior of $\hat q$ as defined by (\ref{eq5}) in a
medium which is moving itself can be found by using simple
arguments based on Lorentz
transformations~\cite{Liu:2006he,Baier:2006pt}
  \be
  \hat q = \gamma_f(1 -
  v_f \cos \th) \, \hat q_0\ ,
  \label{wind}
  \ee
where $v_f $ is the velocity of the wind,
$\gamma_f=1/\sqrt{1-v_f^2}$, and $\theta$ is the angle between the
direction of motion of the hard parton and the direction of the
wind. $\hat q_0$ is the value of $\hat q$ in the absence of a
wind. The result (\ref{wind}) for the dependence of $\hat q$ on
collective flow is valid in QCD and in $\NN=4$ SYM and in the
quark-gluon plasma of any other gauge theory, since its derivation
relies only on properties of Lorentz transformations. If we
crudely guess that head winds are as likely as tail winds, and
that the typical transverse wind velocity seen by a high energy
parton is about half the speed of light, $\hat q$ is increased
relative to that in (\ref{qhat}) by a factor of 1.16. A credible
evaluation of the consequences of (\ref{wind}) for the
time-averaged $\hat q$ extracted from data will, however, require
careful modelling of the geometry of the collision and the
time-development of the collective flow velocity.

It is important to emphasize that the motivation behind
calculating $\hat q$ as defined in~(\ref{eq5}) for $\NN=4$ SYM and
other non-abelian gauge theories is not to understand the full
process of energy loss of a high energy parton in those theories.
Rather, one seeks insights about $\hat q$ in QCD by calculating
the analogous quantity in other theories.

The energy loss of an external heavy quark moving in an $\NN=4$
SYM plasma has also been explored
recently~\cite{Herzog:2006gh,Casalderrey-Solana:2006rq,Gubser:2006nz}.
On general grounds, one expects the momentum of the quark to
satisfy a Langevin equation
 \bea
 \label{lang1}
 {d p_L  \ov dt}  =  - \mu (p_L) p_L + \xi_L (t)\ , \qquad
 %\label{lang2}
 {d p_T  \ov dt}  =   \xi_T (t)\ ,
 \eea
with
 \bea
 \label{lang3}
 \vev{\xi_L (t) \xi_L (t')}  =  \kappa_L (p) \delta (t-t') \,
 \qquad
% \label{lang4}
 \vev{\xi_T (t) \xi_T (t')}  =  \kappa_T (p) \delta (t-t')\ .
 \eea
$p_L$ and $p_T$ are the longitudinal and transverse momentum of
the quark and $\kappa_L (p), \kappa_T (p)$ describe longitudinal
and transverse momentum squared transferred to the quark per unit
time. It was found
in~\cite{Herzog:2006gh,Casalderrey-Solana:2006rq} that the drag
$\mu (p_L)$ is
 \be \label{enrl}
  \mu (p) = {\pi \sqrt{\lam} \ov 2m } T^2 \ .
 \ee
The momentum-independence of the drag in Eq.~(\ref{enrl})
highlights that in the high energy limit the energy loss mechanism
in strongly coupled $\NN=4$ SYM theory is very different from that
in QCD, in which as a result of asymptotic freedom, the dominant
energy loss mechanism is perturbative gluon
radiation\footnote{$\NN=4$ SYM is strongly coupled at all scales,
unlike QCD, which is strongly coupled at scales $\sim T$, but, in
the high (initial quark) energy limit, is weakly coupled at scales
$\sim {\bf k}_{\perp}$, the momentum of typical radiated gluons.}.
  $\kappa_L (p), \kappa_T
(p)$ have also been
computed~\cite{Casalderrey-Solana:2006rq,Gubser:2006nz}
 \be \label{rg}
 \kappa_T = {\pi \sqrt{\lam} \ov (1-v^2)^{1 \ov 4}} T^3\ , \qquad
 \kappa_L = {\pi \sqrt{\lam} \ov (1-v^2)^{5 \ov 4}} T^3 \ .
 \ee
The divergence at $v=1$ for these quantities precludes the use of
$\kappa_T$ as $\hat{q}$ which is defined based on the BDMPS energy
loss formalism, where the initial quark or (radiated gluon) moves
strictly along the light-cone. There is no inconsistency, however,
because for a fixed quark mass $M$, (\ref{rg}) applies only to
velocities $v$ satisfying~\cite{Gubser:2006nz} \be
 {\sqrt{\lam} \ov M} <  { (1-v^2)^{1 \ov 4} \ov T} \
 \label{critv}
 \ee
and $v=1$ singularity in (\ref{rg}) is never reached.
 We will elaborate more on the
physical meaning of (\ref{critv}) in section 4.2.

\section{Quarkonium suppression from velocity scaling of screening length }

The dissociation of charmonium and bottomonium bound states has
been proposed as a signal for the formation of a hot and
deconfined quark-gluon plasma~\cite{Matsui:1986dk}. Recent
analyses of this phenomenon are based on the study of the
quark-antiquark static potential extracted from lattice QCD (see
e.g.~\cite{Satz:2005hx}). In these calculations, the
$q\bar{q}$-dipole is taken to be at rest in the thermal medium. In
heavy ion collisions, however, quarkonium bound states are
produced moving with some velocity $v$ with respect to the medium.
 In the limit of large quark mass, the velocity-dependent dissociation of such a
moving $q\bar{q}$-pair in a medium can be found by  evaluating
$\vev{W^F (\CC_{\rm static}^{\rm boosted})}$ with $\CC_{\rm
static}^{\rm boosted}$ depicted in Fig.~\ref{fig0}. The
orientation of the loop in the $(x_3,t)$-plane changes as a
function of $v$. As a working definition, we may take
 \begin{equation}
    \langle W^F(\CC_{\rm static}^{\rm boosted}) \rangle = \exp \left[ - i\, {\cal T}\,
         E(L) \right] \ .
    \label{2.2}
\end{equation}
where $E(L)$ is (renormalized) free energy of the quark-antiquark
system with self-energy of each quark subtracted. Due to screening
in the medium, one expects $E(L)$ to become flat for large $L$.
The evaluation of (\ref{2.2}) with $ v \neq 0$ in QCD for a
strongly coupled medium is not known at the moment.

From AdS/CFT, one can again evaluate (\ref{2.2}) for $\NN=4$ SYM
plasma in the limit of large $N$ and large $\lam = g_{SYM}^2 N$
using~(\ref{3.7}). %For $v=0$ (i.e. $\CC_{\rm static}$)
One finds that the worldsheet is time-like and thus $S(\CC_{\rm
static})$ is real, giving rise to a real $E(L)$.  Furthermore,
there exists an $L_{max}$ beyond which $E(L)$ becomes identically
zero and thus $L_{max}$ can be interpreted as the screening
length~\cite{Rey:1998bq}. In~\cite{Liu:2006nn,Liu:2006he} (see
also~\cite{Peeters:2006iu}), it was found that %as $v$ increases
%from $0$,
$L_{\rm max}$ changes with $v$ as
 \be
 L_{\rm max} \sim %{f(v,\theta)
  {1\ov T} (1-v^2)^{1 \ov 4} \ .
 \qquad
% \cosh \eta={1 \ov \sqrt{1-v^2}}
 \label{lmax}
 \ee
%where $\th$ is the angle between the orientation of the quark pair
%and the velocity of the moving thermal medium in the rest frame of
%the dipole (in fig.\ref{fig0}, $\th={\pi \ov 2}$).  $f (v, \th)$
%is only weakly dependent on both of its arguments. That is, it is
%close to constant.
%For a given quark mass $M$, (\ref{lmax})
%applies to a velocity $v$ which satisfies~(\ref{critv}). When $v$
%is so close to $1$  that (\ref{critv}) is no longer satisfied, one
%finds that the string worldsheet becomes spacelike and the
%behavior of the Wilson loop is qualitatively similar to that of a
%light-like loop~(\ref{eq5}). We will comment more about this in
%section 4.2.

If the velocity-scaling of $L_{max}$~(\ref{lmax}) holds for QCD,
it will have qualitative consequences for quarkonium suppression
in heavy ion collisions~\cite{Liu:2006nn}.  It implies that the
temperature $T_{\rm diss}$ needed to dissociate the $J/\Psi$
decreases as
 \be \label{rro}
 T_{\rm diss} (v) \sim  T_{\rm diss} (v=0) (1-v^2)^{1/4} \ .
  \ee
   This indicates that $J/\Psi$
suppression at RHIC may increase markedly for $J/\Psi$'s with
transverse momentum $p_T$ above some threshold, on the assumption
that RHIC temperature does not reach the dissociation temperature
$2.1 T_c$ of $J/\Psi$ at zero
velocity~\cite{Karsch:2005nk,Satz:2005hx}. The kinematical range
in which this novel quarkonium suppression mechanism is
operational lies within experimental reach of future
high-luminosity runs at RHIC and will be studied thoroughly at the
LHC in both the $J/\Psi$ and Upsilon channels. We should also
caution that in modelling quarkonium production and suppression
versus $p_T$ in heavy ion collisions, various other effects like
secondary production or formation of $J/\Psi$ mesons outside the
hot medium at high $p_T$~\cite{Karsch:1987zw} remain to be
quantified.  The quantitative importance of these and other
effects may vary significantly, depending on details of their
model implementation. In contrast, Eq. (\ref{rro}) was obtained
directly from a field-theoretic calculation and its implementation
will not introduce additional model-dependent uncertainties.

\section{Discussions}

\subsection{$\NN=4$ SYM versus QCD}

Given that $\hat q$ calculated in $\NN=4$ SYM theory is close to
the value extracted from  RHIC data, is this agreement meaningful
or accidental?  More generally, in what respects can the strongly
interacting plasma of $\NN=4$ SYM theory give a reasonable
description of the quark-gluon plasma in QCD? After all, at a
microscopic level $\NN=4$ SYM is very different from QCD:
\begin{itemize}
 \item The theory is conformal, supersymmetric and contains additional global
symmetry.
 \item The coupling does not run and there is no
confinement.
 \item  No chiral symmetry and no chiral symmetry breaking.
 \item No fundamental quarks,
 additional scalar and fermionic fields in the adjoint representation.
\end{itemize}
These features imply that physics {\it near the vacuum} or {\it at
high energy} are very different between two theories. However, for
QCD at  RHIC temperature (about $2 T_c$) almost all differences
mentioned above become murky and may be irrelevant:

\begin{itemize}

\item There are a variety of indications from lattice QCD
calculations (see e.g.~\cite{Karsch:2006sf} for a review) that QCD
thermodynamics is reasonably well approximated as conformal in a
range of temperatures from about $2T_c$ up to some higher
temperature not currently determined.

\item Supersymmetry of $\NN=4$ SYM is badly broken at finite
temperature for physics of the scale of the temperature.

\item  Above $T_c$ in QCD, there is no confinement and no chiral
condensate. Also if the quark-gluon plasma in QCD is strongly
interacting, as indicated by data from RHIC, the asymptotic
freedom is not important for those physical quantities probing
intrinsic properties of the medium.

\item In a strongly interacting liquid there are, by definition,
no well-defined, long-lived quasiparticles anyway, making it
plausible that observables or ratios of observables can be found
which are insensitive to the differences between microscopic
degrees of freedom and interactions.

\end{itemize}

Thus, it does not seems to be too far-fetched to imagine that the
quark-gluon plasma of QCD, as explored at RHIC and in lattice QCD
calculations, and that of $\NN=4$ SYM share certain common
properties. Indeed the list of similarities between two theories
is growing fast. Examples include the values of thermodynamic
quantities like $\ep/T^4$, $P/T^4$, and the velocity of sound, and
the static screening length between a quark and antiquark at rest,
which can all be compared to lattice QCD calculations, and
dynamical quantities like the shear viscosity, $\hat q$,
$\kappa_T$ which can be compared to inferences drawn from RHIC
data. Perhaps~(\ref{rro}) will be added to this list, once
experiments are done. See sec.6.4 of~\cite{Liu:2006he} for more
details and also~\cite{Buchel:2007vy}.

We are used to the idea that all metals, or all liquids, or all
ferromagnets have common, defining, characteristics even though
they may differ very significantly at a microscopic level. It is
clearly of great interest to understand what are the defining
commonalities of quark-gluon plasmas in different theories, and in
what instances do these commonalities allow qualitative or
semi-quantitative lessons learned about the quark-gluon plasma of
one theory to be applied to that of another.

\subsection{Lightlike versus time-like Wilson loops}

Comparing Eq.~(\ref{2.2}) and Eq.~(\ref{eq5}) we see that while
the exponent in (\ref{2.2}) is pure imaginary, that in (\ref{eq5})
is real. So, if we boost a static Wilson loop to the speed of
light, the qualitative behavior of the Wilson loop should change
as $v \to 1$. How can this happen?

To answer this question, let us first consider a physical set-up
in which a boosted {\it single} Wilson line with $0 < v < 1$ is
realized. The Wilson line can be considered as the non-abelian
phase accumulated along the worldline of an external heavy quark
moving with a constant velocity $v$. In order for a quark to move
at a constant velocity in the medium, some external force has to
be supplied, e.g. by applying an external electric field. However,
as discussed in the second reference in~\cite{Gubser:2006nz}, for
a quark with mass $M$, such a set-up can only be realized for
velocity $v$ not too close to $1$. Otherwise the required electric
field is so large that it will create pairs of quark and
anti-quark. In the context of $\NN=4$ SYM, such a consideration
leads to the inequality~(\ref{critv}). In the limit $M \to
\infty$, the boundary of~(\ref{critv}) is pushed to $v \to 1_-$,
but never exactly $v=1$.

Similarly, %for $v$ not too close to $1$,
one can visualize a boosted Wilson loop with $v < 1$ as the phase
associated with the trajectories of a pair of external heavy quark
and antiquark, with external forces applied to each of them to
keep them moving at a constant velocity $v$. Again, for finite
quark mass $M$, such a set-up can only be realized for $v$
satisfying~(\ref{critv}). In this regime, $E(L)$ in (\ref{2.2}) is
real, from which a quark-antiquark potential can be extracted.
When $v$ is so close to $1$ that~(\ref{critv}) is not satisfied,
instead, a physical way to realize the Wilson loop is to imagine
the quark-antiquark pair as Fock states of a high energy virtual
photon in a deep inelastic scattering experiment. As discussed
in~\cite{Liu:2006ug,Liu:2006he}, unitarity then requires that the
Wilson loop to have a real exponent as in~(\ref{eq5}). In the $M
\to \infty$ limit, the boundary between two regimes lies at $v \to
1_{-} $.
%This expectation is indeed
%realized by the AdS/CFT calculation~\cite{Liu:2006he}. % as noted
%below~(\ref{lmax}).
In~\cite{Liu:2006he}, an alternative interpretation
of~(\ref{critv}) was given: when $v$ does not
satisfies~(\ref{critv}), the Compton wavelength of a quark becomes
greater than the screening length of the medium. From this
perspective, one also sees that the notion of a static quark
potential is not meaningful.

\ack

I would like to thank Krishna Rajagopal and Urs Wiedemann for
collaboration on the work presented here and suggestions on the
draft. I also want to thank Wit Busza for help in preparing the
talk.

\end{document}